\documentclass[12pt]{article}
\usepackage{color}
\usepackage{amssymb}
\usepackage{graphicx}
\usepackage[hang,small,bf]{caption}
\usepackage{hyperref}

\newcommand{\be}[3]{\begin{equation}  \label{#1#2#3}}
\newcommand{\ee}{\end{equation}}
\newcommand{\ba}{\begin{array}}
\newcommand{\ea}{\end{array}}
\newcommand{\bea}[3]{\begin{eqnarray}  \label{#1#2#3}}
\newcommand{\eea}{\end{eqnarray}}

\newcommand{\nn}{\nonumber}

\let\Large=\large
\let\large=\normalsize

\newcommand{\haken}{\mathbin{\hbox to 8pt{%
                 \vrule height0.4pt width7pt depth0pt
                 \kern-.4pt
                 \vrule height4pt width0.4pt depth0pt\hss}}}

\renewcommand{\theequation}{\thesection.\arabic{equation}}

\def\openone{\leavevmode\hbox{\small1\kern-3.8pt\normalsize1}}

\def\e{\epsilon}

\def\bo{{\raise.15ex\hbox{\large$\Box$}}}               
\def\face{{\raise.2ex\hbox{$\displaystyle \bigodot$}\mskip-2.2mu \llap {$\ddot
        \smile$}}}                                      
\def\dg{\dagger}                                     

\def\leftrightarrowfill{$\mathsurround=0pt \mathord\leftarrow \mkern-6mu
        \cleaders\hbox{$\mkern-2mu \mathord- \mkern-2mu$}\hfill
        \mkern-6mu \mathord\rightarrow$}       
\def\dvec#1{\vbox{\ialign{##\crcr
        \leftrightarrowfill\crcr\noalign{\kern-1pt\nointerlineskip}
        $\hfil\displaystyle{#1}\hfil$\crcr}}}           

\def\beq{\begin{equation}}
\def\eeq{\end{equation}}

\def\beqx{\begin{displaymath}}
\def\eeqx{\end{displaymath}}

\def\beqa{\begin{eqnarray}}
\def\eeqa{\end{eqnarray}}

\begin{document}
\DeclareGraphicsExtensions{.jpg,.pdf,.mps,.png}
\begin{flushright}
\baselineskip=12pt
EFI-09-25 \\
MADPH-09-1546\\
\end{flushright}

\begin{center}
\vglue 1.5cm

{\Large \bf Phenomenological Implications of  Supersymmetric Family Non-universal $U(1)^\prime$ Models}
\vglue 2.0cm 
{\Large Lisa L.~Everett$^{a}$, Jing Jiang$^{a}$, Paul G.~Langacker$^{b}$ and Tao Liu$^{c}$ }
\vglue 1cm {
$^a$Department of Physics, University of Wisconsin, \\ Madison, WI 53706\\ \vglue 0.2cm
$^b$School of Natural Science, Institute for Advanced Study, \\
             Einstein Drive, Princeton, NJ 08540
\\ \vglue 0.2cm
$^c$Enrico Fermi Institute, 
University of Chicago, \\5640 S. Ellis Ave., Chicago, IL
60637}
\end{center}

\vglue 1.0cm
\begin{abstract}
We construct a class of anomaly-free supersymmetric $U(1)^\prime$ models that are characterized by family non-universal $U(1)^\prime$ charges motivated from $E_6$ embeddings.   The family non-universality arises from an interchange of the standard roles of the two $SU(5)$ ${\bf 5}^*$ representations within the ${\bf 27}$ of $E_6$ for the third generation.  We analyze $U(1)^\prime$ and electroweak symmetry breaking and present the particle mass spectrum. The models, which include additional Higgs multiplets and exotic quarks at the TeV scale, result in specific patterns of flavor-changing neutral currents in the $b\to s$ transitions that can accommodate the presently observed deviations in this sector from the SM predictions.
\end{abstract}

\bigskip

\section{Introduction}
Extensions of the Standard Model of particle physics (SM) with an additional anomaly-free gauged $U(1)^\prime$ symmetry broken at the TeV scale are arguably some of the most well-motivated candidates for new physics (for a review, see \cite{Langacker:2008yv}).  Such symmetries are theoretically motivated, as they represent the simplest augmentations of the SM gauge sector and are ubiquitous within string and/or grand unified theories.  While the phenomenology of such $Z^\prime$ gauge bosons depends on the details of the couplings of the $Z^\prime$ to the SM fermions,  current limits from direct and indirect searches indicate typical lower bounds of order $800-900$ GeV on the $Z^\prime$ mass and an upper bound of $\sim 10^{-3}$ on the $Z-Z^\prime$ mixing angle \cite{Erler:2009jh}.  For a reasonable range of couplings, the presence of such TeV scale $Z^\prime$ bosons should be easily discernable at present and forthcoming colliders such as the Tevatron and the Large Hadron Collider (LHC). 

Within the context of supersymmetric theories, a plethora of $U(1)^\prime$ models have been proposed, including scenarios motivated by grand unified theories (GUTs) such as $SO(10)$ and $E_6$ and scenarios motivated from string compactifications of heterotic and/or Type II theories (see \cite{Langacker:2008yv} for a review). Recent models also include scenarios in which the $U(1)^\prime$ mediates supersymmetry breaking \cite{Langacker:2007ac}, plays a role in the generation of neutrino masses  \cite{durmus} and/or spontaneous $R$-parity violation \cite{pavel}, or provides a portal to a hidden/secluded sector (for reviews, see~\cite{Langacker:2009im,Goodsell:2009xc}).  Though the details of the $U(1)^\prime$ charge assignments are model-dependent, generically the cancellation of $U(1)^\prime$ anomalies requires an enlargement of the matter content to include SM exotics and SM singlets with nontrivial $U(1)^\prime$ charges.  In these theories, the SM singlets also typically play an important role in triggering the low-scale breaking of the $U(1)^\prime$ gauge symmetry.

In most models of this type, the $U(1)^\prime$ charges of the quarks and leptons are family universal.  Though this feature is desirable for the first and second generations due to the strong constraints from flavor-changing neutral currents (FCNCs),  there is still room for departures from family universality for the charges of the third generation.   In fact, this often occurs in string constructions if the families result from different embeddings (see e.g., \cite{Cleaver:1998gc,Blumenhagen:2005mu}).  Indeed, though many of the results from the $B$ factories have indicated a strong degree of consistency with the Cabibbo-Kobayashi-Maskawa (CKM) predictions of the SM, there are hints of non-SM FCNC patterns within the  $b\to s$ transitions for both $\Delta B=1$ and $\Delta B=2$ processes at the level of a few standard deviations \cite{Bona:2008jn}.
Of the many options for new physics models that can explain this discrepancy, family non-universal $U(1)^\prime$ models are interesting in that they are theoretically well-motivated scenarios that lead to tree-level FCNC, as opposed to scenarios in which the new physics contributions are loop-suppressed~\cite{Langacker:2000ju}.  A recent model-independent analysis of $Z^\prime$-mediated FCNC in the $b\to s$ transitions showed that this general framework can accommodate the data \cite{Barger:2009eq,Barger:2009qs}.  (Related analyses include \cite{zprimefcnc,zprimefcnc2}).  However, it is optimal to consider the bounds on specific family non-universal $U(1)^\prime$ models in addition to the fully model-independent results.

Our purpose in this paper is to construct and analyze supersymmetric anomaly-free family non-universal $U(1)'$ models (which we will denote as NUSSM models).  Our strategy in building this class of NUSSM models is to exploit the well-known fact that in $E_6$ models, there are two options for embedding the down quarks and lepton doublets in the ${\bf 5^*}$ representation of $SU(5)$, which is related to the fact that the down-type Higgs and the lepton doublets have the same gauge quantum numbers.  By choosing one embedding for the first and second generations and the alternative embedding for the third generation, we can obtain anomaly-free models in which the additional family non-universal $U(1)^\prime$ is given by a particular linear combination of the usual $U(1)_\psi$ and $U(1)_\chi$ of $E_6$-inspired models.  

This paper is structured as follows.  We begin by outlining our basic procedure and presenting the resulting classes of anomaly-free family non-universal $U(1)^\prime$ models.  In the following section,  we analyze the gauge symmetry breaking and comment on general features of the mass spectrum.  We next turn to an analysis of the implications of these models for FCNC in the $b\to s $ transitions, then provide our concluding remarks.

\section{$E_6$-Motivated Family Non-universal $U(1)^\prime$ Models (NUSSMs)}
\begin{table}[t]
\caption{The decomposition of the fundamental  ${\bf 27}$
representation of $E_6$ with respect to the $SO(10)$, $SU(5)$, and $U(1)^\prime$ subgroups.  $U(1)_I$ and $U(1)_S$ correspond to specific linear combinations of the $U(1)_{\chi}$ and $U(1)_{\psi}$ gauge groups that result from $SO(10)\rightarrow SU(5)\times U(1)\chi$ and $E_6\rightarrow SO(1)\times U(1)_\psi$, respectively.}
\begin{center}
\resizebox{150mm}{!}{
\begin{tabular}{|c| c| c| c|c| c|c|c|}
\hline $E_6$& $SO(10)$ & $SU(5)$ (1st and 2nd families) & $SU(5)$ (3rd family) & $2 \sqrt{10} Q_{\chi}$ & $2 \sqrt{6}
Q_{\psi}$  & $2Q_I$ & $2\sqrt{15} Q_S$\\
\hline
& {\bf 16}   &   $\Psi_{10} = (u_L,d_L, u^c_L,  e^c_L )$ &$\Phi_{10} =(t_L,b_L, t^c_L,  \tau^c_L )$& $-1$ & 1  & 0 & $-1/2$\\
    &        &   $\Psi_{5^*} = (  d^c_L, \nu_L ,e_L)$ &$\Phi_{5^*} = (\Delta^c, H_d)$  & 3  & 1     & $1$   & $4$  \\
       &     &   $\Psi_1 = \nu_L^c$        &   $\Phi_1 = S$     & $-5$ & 1    &$-1$  &  $-5$ \\
\cline{2-8}
  {\bf 27}   &  {\bf 10}   &   $\sigma_5=(D,h_u)$  &  $\Sigma_5=(\Delta, H_u)$  & 2  & $-2$  &    0 & $1$ \\
        &    &   $\sigma_{5^*}=(D^c, h_d)$ & $\Sigma_{5^*}= (  b^c_L, \nu_{\tau_L}, \tau_L)$ & $-2$ &$-2$ & $-1$& $-7/2$\\
\cline {2-8}
      & {\bf 1}    &   $\sigma_0 = s $        &      $\Sigma_0 = \nu_{\tau_L}^c$     &  0 & 4  & $1$& $5/2$\\
\hline
\end{tabular}}
\end{center}
\label{table1}
\end{table}
In $U(1)^\prime$ models, the cancellation of gauge anomalies generally implies that additional fermions are present in the theory (see e.g.,~\cite{Langacker:2008yv}).  To motivate the presence of these additional fermions and construct simple anomaly-free family non-universal models, our approach is to exploit the properties of $E_6$ embeddings of the SM fermions and Higgs fields in grand unified theories.  Recall that in $E_6$ models, the SM particles are embedded in the fundamental ${\bf 27}$  representations.  With respect to the two-step breaking scheme of $E_6$ to its $SO(10)$ and $SU(5)$ subgroups
\begin{equation}
E_6\rightarrow SO(10)\times U(1)_\psi \rightarrow SU(5)\times U(1)_\chi \times U(1)_\psi,
\end{equation}
the ${\bf 27}$ has the decomposition
\begin{equation}
{\bf 27}= {\bf 16}+{\bf 10}+{\bf 1}=({\bf 10}+{\bf 5^*}+{\bf 1})+({\bf 5}+{\bf 5^*})+{\bf 1},
\end{equation}
with respect to the representations of $SO(10)$ and $SU(5)$, respectively.  Hence, the ${\bf 27}$ has two ${\bf 5^*}$ multiplets; these representations are used to embed the down-type $SU(2)$-singlet quarks with the lepton doublets and exotic $SU(2)$-singlet quarks with down-type Higgs doublets.   A standard choice for model-building is to have the down-type quarks and lepton doublets of all three SM families in the ${\bf 5^*}$ of the ${\bf 16}$, though models with the SM down-type quarks and lepton doublets in the other ${\bf 5^*}$ have also been considered in the literature~\cite{alternative,Athron:2009bs}. 

We will assign the down-type quark singlets and lepton doublets of the first and second generations to be in the ${\bf 5^*}$ of the ${\bf 16}$, and the associated particles of the third generation to be in the  ${\bf 5^*}$ of the ${\bf 10}$, as shown in Table~\ref{table1}.   The matter content of these theories thus includes the following fields: (i) the SM first and second families $\{ \Psi_{10}^i, \Psi_{5^*}^i, \Psi_1^i \}$, Higgs plus exotic fields  $\{\sigma^i_5,\sigma^i_{5^*}\}$, and singlets $\sigma_0^i$ ($i=1,2$ is a family index), and (ii) the SM third family $\{ \Phi_{10}, \Sigma_{5^*}, \Sigma_0\}$, Higgs and exotics $\{ \Sigma_5,\Phi_{5^*} \}$, and singlet $\Phi_1$.  The Higgs sector of the theory thus generically has multiple Higgs doublets and singlets beyond those of the MSSM.\footnote{MSSM-type gauge unification requires the introduction of an additional non-chiral Higgs pair $h+h^*$ from an incomplete ${\bf 27}+{\bf 27}^*$~\cite{Langacker:1998tc}.}

The additional family non-universal $U(1)^\prime$ in these NUSSM models is then a linear combination of the $U(1)_\chi$ and the $U(1)_\psi$ gauge groups:
\begin{eqnarray}
Q^{\prime} = \cos\theta \ Q_{\chi} + \sin\theta \ Q_{\psi}
\label{202}
\end{eqnarray}
(the assumption is that the orthogonal linear combination of $U(1)_\chi$ and $U(1)_\psi$ is either  absent or broken at  a high scale).  
The familiar $U(1)_\eta$ group, which has $\tan\theta=-\sqrt{5/3}$, is family universal and therefore is not useful for our purposes.\footnote{This linear combination occurs in certain Calabi-Yau compactifications of heterotic string theory if $E_6$ breaks to a rank 5 group via the Hosotani mechanism \cite{Witten:1985xc}.}  Two viable options for the additional $U(1)^\prime$ group are: 
\begin{itemize}
\item ${\bf U(1)_I}$ ($\tan\theta=\sqrt{3/5}$).  In this model (the {\it inert} model) the $U(1)'$ gauge boson couplings to the up-type quarks vanish~\cite{Langacker:1984dc} .  Hence, the production of the associated $Z^\prime$ boson is suppressed at hadron colliders.  This is especially the case at the Tevatron, since in high-energy $p\bar{p}$ collisions the $Z'$ production via down quarks is suppressed by an order of magnitude relative to up quarks~\cite{Leike:1998wr}.

\item ${\bf U(1)_S}$ ($\tan\theta=\sqrt{5/27}$). This symmetry is motivated by models with a secluded $U(1)'$ breaking sector and a large supersymmetry breaking $A$-term that have (1) an approximately flat potential that results in an appropriate $Z$--$Z'$ mass hierarchy~\cite{Erler:2002pr}; (2) a strong first order electroweak phase transition and large spontaneous CP-violation, which can result in viable electroweak baryogenesis~\cite{Kang:2004pp}. 

\end{itemize}
While we use the $E_6$ framework to motivate the matter content and $U(1)^\prime$ charges of these models, we do not work within a full grand unified theory.  More precisely, we do not impose the $E_6$ Yukawa coupling relations.  This allows for a TeV-scale $U(1)^\prime$ without the danger of rapid proton decay.\footnote{Detailed studies of $E_6$ theories with broken Yukawa relations can be found in~\cite{Athron:2009bs}.}  

The allowed superpotential terms of NUSSM models (assuming a conserved $R$-parity) are the couplings that are consistent with the SM and $U(1)^\prime$ gauge symmetries.   An inspection of Table~\ref{table1} shows that in the language of the $SU(5)$ decomposition, the usual ${\bf 10\ 5^* 5^*}$ and ${\bf 10\ 10\ 5}$ terms that give rise to quark and lepton Yukawa couplings for all three families (including mixing terms between the third family and the other two families) are allowed by both $U(1)_\chi$ and $U(1)_\psi$.  Similarly, these symmetries allow the generation of Yukawa interactions for the exotic quarks and for the Higgs doublets with the SM singlets ({\it i.e.}, mass terms for the exotic quarks and effective $\mu$ terms for the Higgs fields, which will be of importance for gauge symmetry breaking).   

For simplicity, we assume that only the neutral Higgs bosons from the third family ($H_{u,d}$ and $S$) and one of the first two families ($h_{u,d}$ and $s$) acquire vacuum expectation values (VEVs).\footnote{This is actually without loss of generality by appropriate field redefinitions.}
 In this limit, the Higgs bosons and Higgsinos in the other family  have no mixing at leading order with
the other particles. The mass eigenvalues of these particles are determined by the VEVs of  
${h_{u,d}, s, H_{u,d}, S}$ as well as the Yukawa couplings and soft parameters which are not directly involved in the electroweak symmetry breaking. 
In this article, therefore, we will not discuss them in detail. The relevant superpotential terms are then given by ($I,J=1,2,3$ and $i, j = 1,2$ are family indices):
\begin{eqnarray}
W_Y&=&  (f^{IJ}_{d1} h_d + f^{IJ}_{d2} H_d) Q_L^I d_R^J + (h^{IJ}_1 h_u+ h^{IJ}_2 H_u )Q_L^I u_R^J +\nonumber \\&&
(f^{IJ}_{e1} h_d+ f^{IJ}_{e2} H_d) L^I e_R^J 
  + (y^{IJ}_1 h_u+ y^{IJ}_2H_u) L^I \nu^J_R \label{205a} \\
W_{H} &=&  \lambda_1 sh_dh_u +  \lambda_2 sh_dH_u+\lambda_3 S H_dh_u +\lambda_4 S H_dH_u + \Delta W_H,
 \label{205b}
\end{eqnarray}
in which the Yukawa couplings satisfy the relations $
f_{d1}^{i3}=f_{d1}^{33} \equiv 0$, $f_{d2}^{ij}=f_{d2}^{3i} \equiv 0$,  
$f_{e1}^{3i}=f_{e1}^{33} \equiv 0$, $f_{e2}^{ij}=f_{e2}^{i3} \equiv 0$, 
$y_1^{3i}=y_1^{i3} \equiv 0$, and $y_2^{3i}=y_2^{i3} \equiv 0$.   In Eq.~(\ref{205b}), $\Delta W_H$ represents additional superpotential terms that are consistent with $U(1)_I$ or $U(1)_S$, but explicitly break the orthogonal linear combination of  $U(1)_\chi$ and $U(1)_\psi$. These terms are needed to avoid the appearance of undesirable light axions in the low energy theory (see~\cite{Langacker:2008dq} for a recent discussion).  
For the $U(1)_I$ model, $\Delta W_H$ is a bilinear term:
\begin{eqnarray}
\Delta W_H=\lambda_5 s S. \label{206a}
\end{eqnarray}
Although the coupling $\lambda_5$ in $\Delta W_H$ is dimensionful, there is no associated $\mu$ problem in the traditional sense.  This term is not necessary for $U(1)'$ or electroweak symmetry breaking, so its mass scale need not be connected with the electroweak scale. The Giudice-Masiero mechanism \cite{Giudice:1988yz} therefore can be implemented in both gravity- and gauge-mediated breaking frameworks to produce such a term, even though the $\lambda_5$ in the latter case is typically  small.  In the $U(1)_S$ model,  $\Delta W_H$ consists of the trilinear term
\begin{eqnarray}
\Delta W_H=\lambda_5 ss S. \label{211}
\end{eqnarray}
In what follows, we will focus on the $U(1)_I$ model as a concrete and minimal example, and defer the $U(1)_S$ model for future study.

\section{Particle Mass Spectrum and Gauge Symmetry Breaking}
\label{spectrum}
 The gauge group of the $U(1)_I$ NUSSM model  is given by $SU(3)_c\times SU(2)_L\times U(1)_Y\times U(1)_I$, with gauge couplings $g_3$, $g_2$, $g_1$, and $g'$, respectively.  The matter content, which was presented in Table~\ref{table1}, includes three sets of Higgs fields (one pair of doublets and one singlet per family) and three sets of exotic down-type quarks in addition to the MSSM fields.  As previously discussed, we assume that only two of the three Higgs doublet pairs ($H_{u,d}$ and $h_{u,d}$) acquire vacuum expectation values, and hence focus on the couplings of these Higgs fields only. Restricting to this set of terms, the superpotential for the model is given in Eq.~(\ref{205a}), Eq.~(\ref{205b}), and Eq.~(\ref{206a}).  
  
The tree-level Higgs potential (for the neutral components of the fields) is given by $V=V_F+V_D+V_H$, in which
\begin{eqnarray}
V_F &=& |\lambda_1sh_u^0+\lambda_2sH_u^0|^2 + |\lambda_1sh_d^0+\lambda_3SH_d^0|^2 
+|\lambda_3Sh_u^0+\lambda_4SH_u^0|^2+ |\lambda_2sh_d^0+\lambda_4SH_d^0|^2\nonumber \\&&
+ |\lambda_1h_u^0h_d^0+\lambda_2H_u^0h_d^0+\lambda_5S|^2 
+ |\lambda_3h_u^0H_d^0+\lambda_4H_u^0H_d^0+\lambda_5s|^2~,~ \label{206} 
\end{eqnarray} 
\begin{eqnarray} 
V_D &=& {{G^2}\over 8} \left(|h_u^0|^2 - |h_d^0|^2+|H_u^0|^2 - |H_d^0|^2\right)^2
+\frac{g'^2}{8}\left(-
|h_d^0|^2 + |s|^2 + 
|H_d^0|^2 -|S|^2\right)^2 ~,~\, \label{207}  \nonumber \\
\end{eqnarray} 
\begin{eqnarray} 
V_S &=& m_{h_d}^2 |h_d^0|^2 + m_{h_u}^2 |h_u^0|^2 + m_{H_d}^2 |H_d^0|^2 + m_{H_u}^2 |H_u^0|^2 + m_s^2
|s|^2 + m_S^2
|S|^2  \nonumber\\&&
 +(A_{\lambda_1} \lambda_1 s h_d^0
h_u^0 + A_{\lambda_2} \lambda_2 s h_d^0
H_u^0 + A_{\lambda_3} \lambda_3 S H_d^0
h_u^0 \nonumber \\&& + A_{\lambda_4} \lambda_4 S H_d^0
H_u^0 + B_{\lambda_5} \lambda_5 sS + {\rm H. C.} )~,~\ \label{208}
\end{eqnarray} 
where $G^2=g_1^{2} +g_2^2$.  We also include the one-loop contribution to the potential:
\begin{equation}
\Delta V = \frac{1}{64\pi^2} \mbox{STr} {\cal M}^4 (H_i) \left(\ln
\frac{{\cal M}^2 (H_i)}{\Lambda_{\overline{\rm MS}}^2} -
\frac{3}{2} \right), \label{211b}
\end{equation}
in which $ {\cal M}^2 (H_i)$ denotes the field-dependent mass-squared matrices of the theory,
and $\Lambda_{\overline{\rm MS}}$ is the $\overline{\rm MS}$
renormalization scale.  We will only consider the dominant one-loop contributions that arise from the top quark sector: 
\begin{eqnarray}
\Delta V &=& \frac{3}{32\pi^2} \left[ m_{\tilde{t}_1}^4 (H_i)
\left( \ln \frac{m_{\tilde{t}_1}^2 (H_i)}{\Lambda_{\overline{\rm
MS}}^2} - \frac{3}{2} \right) + m_{\tilde{t}_2}^4 (H_i) \left( \ln
\frac{m_{\tilde{t}_2}^2 (H_i)}{\Lambda_{\overline{\rm MS}}^2} -
\frac{3}{2} \right) \right. \nonumber \\ && \left.  - 2 m_t^4
(H_i) \left( \ln \frac{m_t^2 (H_i)}{\Lambda_{\overline{\rm MS}}^2}
- \frac{3}{2} \right) \right] .
\end{eqnarray}
The Higgs potential  allows for a rich structure of CP-violating effects, including explicit CP violation (for complex couplings) and spontaneous CP violation.  In this work, we will assume that all couplings are real and let the potential parameters satisfy some necessary constraints such that spontaneous CP violation can be avoided.  In this case, the vacuum expectation values of the neutral Higgs components can be taken to be real:
\begin{equation}
\langle h_d^0 \rangle =v_1,\;\;  \langle h_u^0 \rangle =v_2, \;\; \langle H_d^0 \rangle =V_1,\;\; \langle H_u^0\rangle =V_2,\;\; \langle s \rangle =s_1,\;\; \langle S \rangle =s_2.
\end{equation}
Before turning to a numerical analysis, we begin with a general discussion of the particle mass spectrum, starting with the gauge bosons.
The $Z-Z^\prime$ mass-squared matrix is
\begin{eqnarray}
M_{Z-Z'} =\left(\matrix{M_{Z}^2 & M_{Z Z'}^2\cr
M_{Z Z'}^2 &  M_{Z'}^2\cr}\right),~ \,
\end{eqnarray}
in which
\begin{eqnarray}
M_{Z}^2 &=& {G^2\over 2}(v_1^2 +v_2^2+V_1^2 + V_2^2)\equiv {G^2\over 2}v^2,
\nonumber \\  M_{ Z'}^2 &=& 2 g'^2 (Q'^2_{h_d}
v_1^2 + Q'^2_{h_u}v_2^2 + Q'^2_s s_1^2 + Q'^2_{H_d}
V_1^2 + Q'^2_{H_u} V_2^2 + Q'^2_S s_2^2),~\nonumber \\
M_{Z Z'}^2 &=& g'G (  Q'_{h_d} v_1^2 - Q'_{h_u} v_2^2+Q'_{H_d}V_1^2 - Q'_{H_u} V_2^2), 
\end{eqnarray} 
with $v^2=v_1^2 +v_2^2+V_1^2 + V_2^2=(174\, {\rm GeV})^2$. The mass-squared eigenvalues  are
\begin{eqnarray}
M_{Z_1, Z_2}^2 = {1\over 2} \left(M_Z^2 + M_{Z'}^2 \mp 
\sqrt {(M_Z^2-M_{Z'}^2)^2 + 4 M_{Z Z'}^4 } \right),
\end{eqnarray} 
and the $Z-Z'$ mixing angle $\alpha_{Z-Z'}$ is 
\begin{eqnarray}
\alpha_{Z-Z'} = {1\over 2} {\rm arctan} \left({{2 M_{ZZ'}^2}
\over\displaystyle {M_{Z'}^2 - M_Z^2}}\right),
\end{eqnarray} 
which is bounded to be less than a few times $10^{-3}$ (see~\cite{Erler:2009jh} for a recent discussion). 
This typically requires that the singlet vacuum expectation values $s_{1,2} \gg 1$ TeV, resulting in a TeV-scale $Z^\prime$ mass.  The charged gauge boson mass is given as usual by
$M_{W^{\pm}}= g_2 v/\sqrt{2}$.

In the basis $\{ {\tilde B}^{\prime}, \tilde B, \tilde W_3^0,
\tilde h_d^0, \tilde h_u^0, \tilde s, \tilde H_d^0, \tilde H_u^0, \tilde S\}$, the neutralino mass matrix is
\begin{eqnarray}
M_{\tilde \chi^{0}} =\left(\matrix{ M_{\tilde \chi^{0}}(3, 3) &
M_{\tilde \chi^{0}}(3, 6)  \cr M_{\tilde \chi^{0}}(3,6)^T& M_{\tilde \chi^{0}}(6, 6)  \cr}\right),~ \,
\end{eqnarray}
in which
\begin{eqnarray}
M_{\tilde \chi^{0}} (3, 3)= \left(\matrix{M_1^{\prime}
&0&0\cr 0&M_1&0\cr 0&0&M_2\cr}\right) ,\  \nonumber  
\end{eqnarray}
\begin{eqnarray}
M_{\tilde \chi^{0}} (3, 6)= \left(\matrix{\Gamma_{h_d}&\Gamma_{h_u}&\Gamma_{s}&\Gamma_{H_d}&\Gamma_{H_u}&\Gamma_{S}\cr -{1\over
\sqrt 2} g_1 v_1 & {1\over \sqrt 2} g_1 v_2 &0& -{1\over
\sqrt 2} g_1 V_1 & {1\over \sqrt 2} g_1 V_2 &0\cr  {1\over
\sqrt 2} g_2 v_1 & -{1\over \sqrt 2} g_2 v_2&0& {1\over
\sqrt 2} g_2 V_1 & -{1\over \sqrt 2} g_2 V_2&0\cr}\right) ,\, 
\end{eqnarray}
\begin{eqnarray}
M_{\tilde \chi^{0}} (6, 6)= \left(\matrix{0& \lambda_1 s_1&  \lambda_1v_2+  \lambda_2V_2 &0& \lambda_2 s_1 &0
\cr   \lambda_1 s_1 &0&  \lambda_1v_1 &\lambda_3s_2  &0&\lambda_3 V_1
\cr  \lambda_1 v_2+ \lambda_2V_2 &   \lambda_1 v_1 &0& & \lambda_2v_1 & \lambda_5
\cr 0&\lambda_3 s_2& 0&0&  \lambda_4 s_2&\lambda_3 v_2+ \lambda_4 V_2
\cr  \lambda_2 s_1 &0& \lambda_2v_1& \lambda_4 s_2 &0& \lambda_4 V_1
\cr 0&\lambda_3 V_1&  \lambda_5 &\lambda_3 v_2+   \lambda_4 V_2&   \lambda_4 V_1 &0 \cr}\right).\ \nonumber 
\end{eqnarray}
In the above, $\Gamma_{\phi} \equiv \sqrt 2 g' Q_{\phi} 
\langle \phi^*\rangle$, and $M_1^{\prime}$, $M_1$, and $M_2$ are the gaugino mass parameters for
$U(1)^{\prime}$, $U(1)_Y$, and $SU(2)_L$, respectively. 
The chargino mass matrix  is
\begin{eqnarray}
M_{\tilde \chi^{\pm}} =\left(\matrix{M_2 & \frac{g_2}{\sqrt 2}  v_2&  \frac{g_2}{\sqrt 2} V_2
\cr \frac{g_2 }{\sqrt 2}v_1 & \lambda_1 s_1&  \lambda_2 s_1\cr \frac{g_2 }{\sqrt 2}V_1& \lambda_3 s_2&\lambda_4 s_2\cr}\right).
\end{eqnarray}
Since $s_{1,2}\gg v_{1,2}, V_{1,2}$ because of the experimental bounds on $\alpha_{Z-Z'}$, the charginos and neutralinos are typically heavy unless the $\lambda$'s are small or the gaugino masses are light. In the latter situation, the lightest chargino and neutralino will be gaugino-like.

The mass-squared matrices of the sfermions (denoted collectively as $\phi$) are
\begin{eqnarray}
M_{\phi}^2 =\left(\matrix{ (M_{\phi}^2)_{11}& (M_{\phi}^2)_{12}\cr (M_{\phi}^2)_{21}&
(M_{\phi}^2)_{22} \cr}\right).
\end{eqnarray}
With the definitions
\begin{eqnarray}
\Delta_{\phi} &\equiv& \frac{G^2}{2}(T_3^{\phi} - Q_{EM}^{\phi} \sin^2\theta_W)(v_1^2-v_2^2+V_1^2-V_2^2)~,~\, \\
\Delta'_{\phi} &\equiv& Q'_{\phi} g'^2 ( Q'_{h_d}v_1^2 + Q'_{h_u} v_2^2 + Q'_s s_1^2+ Q'_{H_d}
V_1^2 + Q'_{H_u} V_2^2  + Q'_S s_2^2 )~,~\  
\end{eqnarray}
the entries for example of the up-type squark mass-squared matrix are:
\begin{eqnarray}
(M_{\tilde u}^2)_{11}& =&m^2_{{\tilde Q}_L} + m^2_{u} +
\Delta_{{\tilde u}_L} + \Delta'_{{\tilde u}_L} ~,~\, \nonumber \\
(M_{\tilde u}^2)_{12}& = & h_1(\lambda_1v_1s_1 + \lambda_3V_1 s_2 ) +h_2(\lambda_2 v_1 s_1 + \lambda_4 V_1 s_2 ) 
-(A_{h_1}h_1v_2 + A_{h_2} h_2 V_2) \nonumber \\
(M_{\tilde u}^2)_{21}& = & (M_{\tilde u}^2)_{12}\nonumber \\
(M_{\tilde u}^2)_{22} &=&m^2_{{\tilde u}_R} + m^2_{u} +
\Delta_{{\tilde u}_R} + \Delta'_{{\tilde u}_R}.~\,
\end{eqnarray}
Analogous expressions can be written for the down-type squarks, sleptons, and sneutrinos.  
The physical Higgs spectrum  consists of 6 CP-even neutral Higgs bosons, 4 CP-odd neutral Higgs bosons, and 6 charged Higgs bosons (not including the second family). The tree-level charged Higgs boson mass-squared matrix is given in the Appendix. 

Next we turn to a numerical analysis of this sector of the model, taking into account the constraints on the $Z'$ gauge boson.  
We explore the viable regions of parameter space in which (i)  $s_1,s_2  \gg v_{1,2}, V_{1,2} $, which is needed for a TeV scale $Z'$, and (ii) $V_2 > V_1 > v_{1,2}$, which is  motivated by the observed hierarchies in the SM fermion mass spectrum.  To obtain an acceptable minimum, typically we need the Higgs soft mass parameters to satisfy $m_s^2$ or $m_S^2 \ll m_{h_u}^2, m_{H_u}^2 < m_{h_d}^2, m_{H_d}^2$. We also set the $U(1)_I$ gauge coupling to $g'=\sqrt{\frac{5}{3}} g_1$ and enforce the following constraints on the Yukawa couplings:\footnote{This approximation must be relaxed slightly to obtain CKM mixing of the third family with the first and second families, but that is irrelevant for our present purposes.}
\begin{eqnarray}
h_1^{33}=h_2^{33} , \qquad h_1^{3i} = h_1^{i3}= h_2^{3i} = h_2^{i3} = 0.
\end{eqnarray}
which result in the condition
\begin{eqnarray}
h_1^{33}=h_2^{33} =  \frac{165\, {\rm GeV}}{v_2+V_2},
\end{eqnarray}
in which we have included the one-loop QCD corrections to the top quark mass.
 
We consider one typical numerical example; the relevant input parameters and results are summarized in Tables~\ref{table2}--\ref{table4}.  The mass spectrum of the neutral Higgs bosons are calculated at one-loop level, and the mass spectra of the other particles are calculated at tree-level.  As a check, we estimate the $Z'$ mass and the $Z-Z'$ mixing angle by using the Higgs VEVs given in Table~\ref{table2}, as follows:
\begin{eqnarray}
M_{Z_2} \approx  M_{Z'} \approx \sqrt{0.18(s_1^2 +s_2^2)} \sim 1.9\,  {\rm TeV},\;\;
\alpha_{Z-Z'} \approx \frac{M^2_{Z-Z'}}{M_{Z'}^2} \sim 0.0003,  
\end{eqnarray}
which is consistent with the detailed results.  The lightest CP-even Higgs boson $H_1$ is a linear combination of the real parts of the four Higgs doublets, with a negligible singlet admixture; the orthogonal linear combinations of these four states are the $H_3$, $H_4$, and $H_6$ bosons.   These heavier Higgs bosons fall into $SU(2)$ multiplets together with the three heaviest CP-odd states and the set of charged Higgs bosons, as follows: $(H_3, A_2,  H_1^\pm)$, $(H_4, A_3,  H_2^\pm)$, and $(H_6, A_4,  H_3^\pm)$.  The second lightest and second heaviest CP-even states are admixtures of the two singlet Higgs fields, as is the lightest CP-odd boson (which has a mass that controlled by the Higgs bilinear terms).    The chargino and neutralino mass spectrum is highly model-dependent, as it is sensitive to the electroweak and hypercharge gaugino masses, which do not strongly impact the gauge symmetry breaking.    Hence, the physics of the lightest superparticle (LSP) can vary greatly depending on the exact structure of the gaugino sector, though the gauge and Higgs sectors can remain almost the same in this case.  In our numerical example, in which the gaugino masses are light and obey GUT relations, the LSP is a predominantly bino-like neutralino that can be an acceptable dark matter candidate in regions of the parameter space.\footnote{The neutralino sector has additional complications due to the presence of the additional Higgs supermultiplets that do not participate in electroweak symmetry breaking at tree level. Hence, a detailed numerical analysis would be needed to ascertain whether the neutralino LSP satisfies the dark matter constraints.  As this is tangential to the main purpose of our paper, we do not address it here.}

Finally, we comment on the exotic colored particles  $\{D^i, D^{ci}\}$ and $\{\Delta, \Delta^c\}$. The exotic scalars do not 
obtain VEVs. They and their superpartners influence the gauge symmetry breaking only at loop level. These exotic particles are chiral, so their tree-level masses can be produced only through Yukawa interactions. The superpotential terms that describe their interactions with the Higgs fields and the corresponding soft supersymmetry breaking terms are 
\begin{eqnarray}
W_{E} &=& \tilde  \lambda_1^{ij} s D^{ci} D^j + \tilde  \lambda_2^i s D^{ci} \Delta+ \tilde \lambda_3^i S \Delta^c D^i + \tilde \lambda_4 S \Delta^c \Delta,  \\
V_{E} &=& A_{\tilde  \lambda_1^{ij}} \tilde  \lambda_1^{ij} s \tilde{D}^{ci} \tilde{D}^j + A_{\tilde  \lambda_2^i} \tilde  \lambda_2^i s \tilde{D}^{ci} \tilde{\Delta}+ A_{\tilde  \lambda_3^i} \tilde \lambda_3^i S \tilde{\Delta}^c \tilde{D}^i + A_{\tilde  \lambda_4}\tilde \lambda_4 S \tilde{\Delta}^c \tilde{\Delta}.
\end{eqnarray}
It is straightforward to determine the mass matrices for these states for given Yukawa couplings and $A$ parameter values.  For ${\mathcal O} (\tilde \lambda_{1,2,3,4}) \sim 0.1$ (where their contributions to the effective neutral Higgs potential can be neglected) 
and not large $A_{\tilde \lambda}$ values, the exotic particles will typically obtain masses of the order of several hundred GeV. 

\begin{table}
\caption{Parameter values and Higgs VEVs. The dimensional parameter values are given in GeV or GeV$^2$. The Higgs VEVs are given in GeV.}
\begin{center}
\begin{tabular}{|c|c|c|c|c|c|}
\hline
$\lambda_1$ & $\lambda_2$&$\lambda_3$&$\lambda_4$&$\lambda_5$& $g'$\\
\hline 0.10 & 0.30 &0.10& $-0.45$ &449&0.60
\\ \hline
$M_1'$&$M_1$&$M_2$& $M_3$ &$m_{\tilde Q_3}^2$& $m_{\tilde T_R}^2$
\\ \hline
 112 &        112 &        224 &        673 &     $1.21 \times 10^5$ &     $1.21 \times 10^5$
\\ \hline
$m_{h_d}^2$& $m_{h_u}^2$& $m_s^2$&$m_{H_d}^2$&$m_{H_u}^2$& $m_S^2$\\
\hline
$9.06 \times 10^5$ &     $7.04 \times 10^5$ &    $1.21\times 10^6$ &     $9.06 \times 10^5$ &   $-1.01 \times 10^6$ &   $-1.21 \times 10^6$
\\ \hline
$A_{\lambda_1}$ & $A_{\lambda_2}$&$A_{\lambda_3}$&$A_{\lambda_4}$&$B_{\lambda_5}$& $A_{\tilde T}$ \\
\hline
$-1350$ &     $ -1350$ &       $-449$ &       1080 &       $-359$ &        897
\\ \hline
$v_1$ & $v_2$ & $s_1$ & $V_1$& $V_2$& $s_2$
\\ \hline
 54.8 &         83.7 &       2000 &         93.3 &        108 &       4010
\\ \hline
\end{tabular} 
\end{center} \label{table2}
\end{table}

\begin{table}
\caption{The particle mass spectrum and $Z-Z'$ mixing angle of the NUSSM model (all masses are in GeV).}
\begin{center}
\begin{tabular}{|c|c|c|c|c|c|c|}
\hline
$M_{Z_2}$& $\sin\theta_{Z-Z'}$ & $m_{\tilde t_1}$ & $m_{\tilde t_2}$  & $m_{\tilde \chi_1^\pm}$ & $m_{\tilde \chi_1^0}$ & / \\
\hline
 1900 & $3.10\times 10^{-4}$ &  275 &        609 & 219 & 114& / \\ \hline
$m_{H_1}$ & $m_{H_2}$ & $m_{H_3}$ & $m_{H_4}$  & $m_{H_5}$  & $m_{H_6}$  & / \\
\hline
  173 &            369 &    1100 &      1640 &       1970 &   2360 &   /   \\ \hline
$m_{A_1}$ & $m_{A_2}$ & $m_{A_3}$ & $m_{A_4}$  & $m_{H_1^\pm}$ & $m_{H_2^\pm}$ & $m_{H_3^\pm}$ \\
\hline
  633 &       1080 &         1650 &   2340 &     1060 &          1630  & 2330 \\
\hline
\end{tabular} 
\end{center}\label{table3}
\end{table}

\begin{table}
\caption{The composition of the neutral Higgs mass eigenstates at the one-loop level.}
\begin{center}
\begin{tabular}{|c|c|c|c|c|c|c|}
\hline
&$h_{dr}^0$& $h_{ur}^0$&$s_{r}$&$H_{dr}^0$&$H_{ur}^0$&$S_{r}$\\
\hline
$H_1$&
        0.31 &
        0.48 &
       -0.09 &
        0.53 &
        0.61 &
       -0.07 
\\
\hline
$H_2$&
   0.07 &
        0.06 &
        0.87 &
        0.06 &
        0.05 &
        0.49
\\
\hline
$H_3$&
    -0.28 &
        0.86 &
       -0.02 &
       -0.34 &
       -0.25 &
        0.03 
\\
\hline
$H_4$&
  -0.88 &
       -0.11 &
        0.02 &
        0.13 &
        0.43 &
        0.04
\\
\hline
$H_5$&
      0.03 &
       -0.01 &
       -0.49 &
        0.04 &
       -0.01 &
        0.87 
\\
\hline
$H_6$&
   -0.20 &
        0.06 &
        0.01 &
        0.76 &
       -0.61 &
       -0.03
\\
\hline
\hline
&$h_{di}^0$& $h_{ui}^0$&$s_{i}$&$H_{di}^0$&$H_{ui}^0$&$S_{i}$\\
\hline
$A_1$&
      -0.04 &
       -0.02 &
        0.89 &
       -0.01 &
       -0.01 &
        0.45
\\
\hline
$A_2$&
     0.26 &
        0.87 &
        0.03 &
        0.35 &
       -0.24 &
        0.02\\
\hline
$A_3$&
    0.89 &
       -0.10 &
        0.04 &
       -0.12 &
        0.43 &
        0.01
\\
\hline
$A_4$&
    -0.20 &
       -0.08 &
       -0.01 &
        0.76 &
        0.62 &
        0.02 
\\
\hline
$G_1$ &
  0.30&
 -0.33&
  0.20&
  0.59&
 -0.52&
 -0.39
\\
\hline
$G_2$&
 -0.12 &
        0.21 &
       -0.40 &
       -0.25 &
        0.26 &
        0.81 \\
\hline
\end{tabular}
\end{center} \label{table4}
\end{table}

\begin{table}
\caption{The composition of the charged Higgs mass eigenstates at tree level.}
\begin{center}
\begin{tabular}{|c|c|c|c|c|}
\hline
&$h_d^-$& $h_u^{+*}$&$H_d^-$& $H_u^{+*}$\\
\hline
$H_1^-$&
 0.26 &
        0.87 &
        0.34 &
       -0.23\\
\hline
$H_2^-$&
   0.89 &
       -0.10 &
       -0.12 &
        0.42
\\
\hline
$H_3^-$&
 -0.20 &
       -0.07 &
        0.75 &
        0.62
\\ \hline
$G_1^-$&      0.31 &
       -0.47 &
        0.55 &
       -0.62\\
\hline
\end{tabular}
\end{center} \label{table5}
\end{table}

\section{$Z^\prime$-mediated FCNC Effects}

\label{FCNC}
In this section, we analyze the $Z'$-induced FCNC effects  After a brief review of the formalism, we will show the results of our correlated analysis of the $\Delta B =1, 2$ processes via $b\to s$ transitions and discuss the resulting parameter space constraints.  
 The processes of interest include $B_s - \bar B_s$ mixing and the time-dependent  CP asymmetries of the penguin-dominated neutral $B_d \to (\phi, \eta', \pi, \rho, \omega, f_0)K_S$ decays.

The FCNC effects in general NUSSM models include both $Z^\prime$-mediated FCNC processes and contributions to FCNC from the soft supersymmetry breaking parameters.  In this work, we assume for simplicity that the soft terms do not result in large FCNC effects (this can be easily achieved; 
see e.g.~\cite{Martin:1997ns}) and consider only the $Z^\prime$ contributions.   We now briefly discuss the  formalism for addressing such $Z^\prime$ effects in the NUSSM (for a model-independent discussion, see~\cite{Langacker:2000ju,Barger:2009eq,Barger:2009qs}).

For the SM fermions $\psi_{L,R}$ with $U(1)^\prime$ charges $\tilde \e_{\psi_{L,R}}$, the fermion mass matrices are diagonalized by the biunitary transformation
$M_{\psi,{\rm diag}}=V_{\psi_R} M_{\psi} V_{\psi_L}^{\dg}$
(the CKM matrix is 
$  V_{\rm CKM} = 
    V_{u_L} V_{d_L}^{\dg}$). 
The chiral $Z'$ couplings in the fermion mass eigenstate basis are
\begin{eqnarray}
  B^{\psi_L}\equiv
    V_{\psi_L}\tilde \e^{\psi_L} V_{\psi_L}^{\dg}\;,
    \qquad
  B^{\psi_R} \equiv
    V_{\psi_R} \tilde \e^{\psi_R} V_{\psi_R}^{\dg}\;.
  \label{453}
\end{eqnarray}
In our $U(1)_I$ model, the only SM fields with nontrivial $U(1)^\prime$ charges are the down-type quark singlets and the lepton doublets: 
\begin{eqnarray}
\tilde \e^{d_R}=-\frac{1}{2}\begin{pmatrix} {1 &0&0 \cr 0 &  1  & 0 \cr 0 & 0&  -1 } \end{pmatrix},\;\;\;
\tilde \e^{L_L}&=&\frac{1}{2}\begin{pmatrix} {1 &0&0 \cr 0 &  1  & 0 \cr 0 & 0&  -1 } \end{pmatrix}.
 \label{454}
\end{eqnarray}
With the unitary matrices $V_{d_R,L_L}$ written as 
\begin{eqnarray}
 V_{d_R,L_L} &=&\begin{pmatrix} { W_{d_R,L_L} & X_{d_R,L_L} \cr  Y_{d_R,L_L} & Z_{d_R,L_L} } \end{pmatrix}, \label{455}
\end{eqnarray}
where $W_{d_R,L_L}$ is a $2\times 2$ submatrix, one obtains 
\begin{eqnarray}
B^{d_R} 
&=&-\frac{1}{2}\begin{pmatrix} { W_{d_R}^\dagger W_{d_R} - Y_{d_R}^\dagger Y_{d_R}  &  W_{d_R}^\dagger X_{d_R} -Y_{d_R}^\dagger Z_{d_R} \cr X_{d_R}^\dagger W_{d_R} - Z_{d_R}^\dagger Y_{d_R} & X_{d_R}^\dagger X_{d_R} - Z_{d_R}^\dagger Z_{d_R} } \end{pmatrix},\nonumber \\
B^{L_L} 
&=&\frac{1}{2}\begin{pmatrix} { W_{L_L}^\dagger W_{L_L} - Y_{L_L}^\dagger Y_{L_L}  &  W_{L_L}^\dagger X_{L_L} -Y_{L_L}^\dagger Z_{L_L} \cr X_{L_L}^\dagger W_{L_L} - Z_{L_L}^\dagger Y_{L_L} & X_{L_L}^\dagger X_{L_L} - Z_{L_L}^\dagger Z_{L_L} } \end{pmatrix}. \label{456}
\end{eqnarray}
To avoid the constraints on non-universality for the first two families from $K- \bar K$ mixing and $\mu-e$ conversion in muonic atoms, we assume small fermion mixing angles or small $X_{d_R,L_L}$, $Y_{d_R,L_L}$ elements. The $Z'$ couplings then take the form
\begin{eqnarray}
 && B^{d_R}_{11}, B^{d_R}_{22} \approx -\frac{1}{2},   \ \ B^{d_R}_{33} \approx \frac{1}{2} , \ \   B^{d_R}_{13}, B^{d_R}_{23} \sim {\mathcal O} (X_{d_R},Y_{d_R}),  \nonumber \\
&& B^{L_L}_{11}, B^{L_L}_{22} \approx  \frac{1}{2},   \ \ B^{L_L}_{33} \approx -\frac{1}{2} , \ \   B^{L_L}_{13}, B^{L_L}_{23} \sim {\mathcal O} (X_{L_L},Y_{L_L}).\label{457}
\end{eqnarray}
Here $B^{d_R,L_L}_{13}$ and $B^{d_R,L_L}_{23}$ generically are complex. 
The $Z'$-induced corrections to the Wilson coefficients in the $U(1)_I$ model\footnote{In the $U(1)'_S$ model, in which all SM fermions are charged under the $U(1)'$ symmetry, the $Z'$-induced corrections to the Wilson coefficients take a more general form (see e.g.~\cite{Barger:2009eq,Barger:2009qs}). We will not discuss them here.} are given by (for the associated operators, see e.g.~\cite{Barger:2009qs}):
 \begin{eqnarray}
 \Delta \tilde C^{B_s}_1& =& - (B_{bs}^R)^2, \;\;
\Delta \tilde C_{3} = - \frac{4}{3 V_{tb} V_{ts}^*}  B_{bs}^R  B_{dd}^R, \;\; 
\Delta \tilde C_{9} = \frac{4}{3 V_{tb} V_{ts}^*}  B_{bs}^R B_{d d}^R ,
 \nonumber \\
\Delta \tilde C_{9V} &=& -\Delta \tilde C_{10A}  = - \frac{2}{V_{tb} V_{ts}^*} B_{bs}^RB_{ll}^L.
 \label{458} 
\end{eqnarray}    
%
To achieve sufficient precision, we need to have an accurate knowledge of the relevant Wilson coefficients at  the $b$ quark mass scale $m_b=4.2$ GeV (for general discussions, see e.g.~\cite{Buchalla:1995vs}). The parameter values used in our calculations are summarized in~\ref{Parameters}.\\

\noindent $\bullet$ {\bf $B_s-\bar B_s$ mixing.} The new physics (NP) contributions to the off-diagonal mixing matrix element are parametrized as
\begin{eqnarray}
M_{12}^{B_s}=(M_{12}^{B_s})_{\rm SM} C_{B_s} e^{2 i \phi_{B_s}^{\rm NP}}, \label{402}
\end{eqnarray}
where $C_{B_s}=1$ and $\phi_{B_s}^{\rm NP}=0$ in the SM limit. Although the data indicate that $C_{B_s}\simeq 1$, a recent analysis~\cite{Bona:2008jn} suggests that $\phi_{B_s}^{\rm NP}$ deviates from zero at the $2-3 \sigma$ level (see Table~\ref{table6});  an earlier discussion was given in~\cite{Lenz:2006hd}. The analysis of \cite{Bona:2008jn} includes all available results on $B_s$  mixing, including the tagged analyses of $B_s \to \psi \phi$ by CDF~\cite{Aaltonen:2007he} and D$\emptyset$~\cite{:2008fj}. As discussed for example in~\cite{Tarantino:2008pb}, this discrepancy disfavors scenarios with minimal flavor violation (MFV), though no single measurement yet has a $3\sigma$ significance.

In our $U(1)_I$ NUSSM model, $C_{B_s}$ and $\phi_{B_s}^{\rm NP}$ at the $m_b$ scale are given by 
\begin{eqnarray}
C_{B_s} e^{2i\phi_{B_s}} &=& 1 - 3.59 \times 10^5  (\Delta C_1^{B_s} + \Delta \tilde C_1^{B_s}) 
+ 2.04 \times 10^6 \Delta \tilde C_3^{B_s}. \label{403}
\end{eqnarray}
The large coefficients of the correction terms in Eq.~(\ref{403}) are due to the fact that the NP is introduced at tree-level while the SM limit is a loop-level effect.\\

\begin{table}[th]
\begin{center}
\begin{tabular}{@{}ccc}
Observable  & $1 \sigma$ C.L.& $2 \sigma$ C.L.  \\
\hline
\hline
$\phi_{B_s}^{\rm NP} [^\circ]$  (S1)            & -20.3 $\pm$ 5.3 & [-30.5,-9.9] \\
    $\phi_{B_s}^{\rm NP} [^\circ]$                         (S2)       & -68.0 $\pm$ 4.8 & [-77.8,-58.2] \\
$C_{B_s}$                           & 1.00 $\pm$ 0.20 & [0.68,1.51] \\
\hline
\hline
\end{tabular}
\end{center}
\caption {The fit results for the $B_s - \bar B_s$ mixing parameters~\cite{Bona:2008jn}. The two $\phi_{B_s}^{\rm NP}$ solutions (``S1'' and ``S2'') result from measurement ambiguities; see~\cite{Bona:2008jn} for details.}
\label{table6}
\end{table}



\noindent $\bullet$ {\bf $B_d \to (\psi,\pi, \phi, \eta', \rho, \omega, f^0)K_S$ decays.}
\begin{table}[t]
\begin{center}
\begin{tabular}{|c|c|c|}
  \hline
$f_{CP}$ & $-\eta_{CP} {\mathcal S}_{f_{CP}}$ (1$\sigma$ C.L.) & ${\mathcal C}_{f_{CP}}$(1$\sigma$ C.L.)   \\  \hline
$\psi K_S$  & $+0.672\pm0.024 $ & $+0.005\pm0.019 $   \\ \hline 
$\phi K_S$ & $+0.44^{+0.17}_{-0.18} $ & $-0.23\pm0.15 $  \\
$\eta^\prime K_S$ & $+0.59\pm0.07  $ & $-0.05\pm0.05$  \\  
$\pi K_S$ & $+0.57\pm0.17$ & $+0.01\pm0.10$  \\
$\rho K_S$ & $+0.63^{+0.17}_{-0.21}$ & $-0.01\pm0.20 $  \\ 
$\omega K_S$ & $+0.45\pm0.24$ & $-0.32\pm0.17 $  \\ 
$f_0 K_S$ & $+0.62^{+0.11}_{-0.13}$ & $0.10\pm0.13$ \\ \hline
\end{tabular}
\end{center}
\caption{The world averages of the experimental results for the CP asymmetries in $B_d$ decays via $b\to\bar qq s$ transitions~\cite{Barberio:2008fa}.} \label{table7}
\end{table}
The direct and the mixing-induced CP asymmetries in hadronic $B_d$ decays are parametrized as follows: 
\begin{eqnarray}
{\mathcal C}_{f_{CP}}
= \frac{1-|\lambda_{f_{CP}}|^2}{1+|\lambda_{f_{CP}}|^2} ~,
\qquad
{\mathcal S}_{f_{CP}}
= \frac{2{\rm Im} \left[ \lambda_{f_{CP}} \right]}{1+|\lambda_{f_{CP}}|^2}.    \label{404}
\end{eqnarray}
in which
\begin{eqnarray}
\lambda_{f_{CP}}
\equiv \eta_{f_{CP}} e^{-2i\phi_{B_d}}
       \frac{\bar{A}_{f_{CP}}}{A_{f_{CP}}}.  \label{405}
\end{eqnarray}
Here $\phi_{B_d}$ is the $B_d-\bar B_d$ mixing angle, $A_{f_{CP}}$ is the decay amplitude of $B_d \to f_{CP}$ ($\bar A_{f_{CP}}$ is its CP conjugate), and 
$\eta_{f_{CP}}=\pm1$ is the CP eigenvalue for the final state $f_{CP}$. In the SM, $\phi_{B_d} = \beta \equiv \arg\left[-(V_{cd}V_{cb}^*)/(V_{td}V_{tb}^*)\right]$, and a non-trivial weak phase enters $A_{f_{CP}} $ only at ${\mathcal O}(\lambda^2)$.  This implies the following SM relation between the decays proceeding via $b\to s \bar qq(q=u,d,c,s)$ and the penguin-dominated modes such as $B_d \to (\pi,\phi, \eta', \pi, \rho, $ $\omega, f^0 ) K_S$:
\begin{eqnarray}
-\eta_{f_{CP}}{\mathcal S}_{f_{CP}} = \sin 2 \beta + {\mathcal O}(\lambda^2),  \ \ \ \  {\mathcal C}_{f_{CP}} = 0 + {\mathcal O}(\lambda^2).   \label{416}
\end{eqnarray}
However, the experimental values of $\sin 2 \beta$ obtained from the penguin-dominated modes are below the SM prediction and the results from the charmed $B_d \to \psi K_S$ mode.  The central values of the direct CP asymmetries of $B_d\to \phi K_S$ and $B_d \to \omega K_S$ are also small  compared to the $B_d \to \psi K_S$ mode, as shown in Table~\ref{table7}.  Since $B_d\to \psi K_S$ is a tree-level process in the SM, large values for $\Delta {\mathcal S}_{f_{CP}}=-\eta_{f_{CP}}{\mathcal S}_{f_{CP}}+ \eta_{\psi K_S}{\mathcal S}_{\psi K_S}$ and $\Delta {\mathcal C}_{f_{CP}}={\mathcal C}_{f_{CP}} - {\mathcal C}_{\psi K_S}$ may indicate the presence of NP in the $b\to s$ transitions.

In NUSSM models, $Z'$-induced FCNC effects can provide dramatic changes to the results, since a new weak phase can enter $A_{f_{CP}}$ at tree level. Following \cite{Ali:1998eb},  the 
$\lambda_{f_{CP}}$ parameters of $B_d \to (\psi, \phi, \eta', \pi, \rho, \omega, f^0)K_S$ at the $m_b$ scale are given by
\begin{eqnarray} 
\lambda_{\psi K_S} &=&(-0.63+ 0.74 i)   \label{406}\\ \nn && [1 -
  (2.93 -2.61 i )(\Delta C_3 + \Delta \tilde C_3)^{*} - (2.94 -2.95
  i)(\Delta C_5 + \Delta \tilde C_5)^{*}\nonumber \\ && +
  (0.18 - 0.01 i )(\Delta C_7 + \Delta \tilde C_7)^{*} -
  (0.06 - 0.04 
  i)(\Delta C_9 + \Delta \tilde C_9)^{*}]\nonumber \\ &&/[ 1 - (2.80+2.61 i
  )(\Delta C_3 + \Delta \tilde C_3) - ( 2.74+2.99 i)(\Delta C_5 +
  \Delta \tilde C_5)\nonumber \\ &&+ (0.17 + 0.01 i
  )(\Delta C_7 + \Delta \tilde C_7) - (0.04 + 0.05 i)(\Delta C_9 +
  \Delta \tilde C_9)]~,  \nonumber 
\end{eqnarray} 
\begin{eqnarray} 
\lambda_{\pi K_S} &=&(-0.70+ 0.70 i)   \label{407}\\ \nn && [1 -
  (1.09 +0.50 i )(\Delta C_3 + \Delta \tilde C_3)^{*} - (6.73 +2.79
  i)(\Delta C_5 + \Delta \tilde C_5)^{*}\nonumber \\ && -
  (9.68 +3.21 i )(\Delta C_7 + \Delta \tilde C_7)^{*} +
  (13.86 +4.48 
  i)(\Delta C_9 + \Delta \tilde C_9)^{*}]\nonumber \\ &&/[ 1 + (1.08+0.48 i
  )(\Delta C_3 + \Delta \tilde C_3) - ( 6.66+2.70 i)(\Delta C_5 +
  \Delta \tilde C_5)\nonumber \\ &&- (9.58 + 3.09 i
  )(\Delta C_7 + \Delta \tilde C_7) +(13.71 + 4.31 i)(\Delta C_9 +
  \Delta \tilde C_9)]~,  \nonumber 
\end{eqnarray} 
\begin{eqnarray} 
\lambda_{\phi K_S} &=& (-0.70 + 0.70 i)  \label{408}\\ \nn && [1  -
  ( 28.62+11.37i )(\Delta C_3 + \Delta \tilde C_3)^{*} - (24.08 + 10.41
  i)(\Delta C_5 + \Delta \tilde C_5)^{*}\nonumber \\ &&+
 (14.57 + 5.88 i) (\Delta C_7 + \Delta \tilde C_7)^{*} + (15.08 +
 5.92 i) (\Delta C_9 + \Delta \tilde C_9)^{*}]\nonumber \\ &&/[  1 - (28.27+10.89 i
  )(\Delta C_3 + \Delta \tilde C_3) - ( 23.80+10.00 i)(\Delta C_5 +
  \Delta \tilde C_5)\nonumber \\ &&+ (14.39 + 5.64 i)
 (\Delta C_7 + \Delta \tilde 
        C_7) + (14.90 + 5.67 i)
        (\Delta C_9 + \Delta \tilde C_9)]~,  \nonumber 
\end{eqnarray} 
\begin{eqnarray} 
\lambda_{\eta^\prime K_S} &=& (-0.70 + 0.69 i)   \label{409}\\ \nn 
&& [1  -
  (10.88+3.29 i )(\Delta C_3 + \Delta \tilde C_3)^{*} +(8.26+2.06
  i)(\Delta C_5 + \Delta \tilde C_5)^{*}\nonumber \\ &&+
  (2.11 + 0.67 i)(\Delta C_7 + \Delta \tilde C_7)^{*} + (2.10 +
  0.54 i)(\Delta C_9 + \Delta \tilde C_9)^{*}]\nonumber \\ &&/[ 1 - (10.73+3.21 i
  )(\Delta C_3 + \Delta \tilde C_3) + ( 8.14+2.00 i)(\Delta C_5 +
  \Delta \tilde C_5)\nonumber \\ &&+ (2.08 + 0.65
  i)(\Delta C_7 + \Delta \tilde C_7) + (2.07 + 0.52 i)(\Delta C_9 +
  \Delta \tilde C_9)]~,  \nonumber 
\end{eqnarray}
\begin{eqnarray} 
\lambda_{\rho K_S} &=& (-0.74 + 0.65 i)  \label{411}\\ \nn && [1  +
  (0.26 + 0.06 i )(\Delta C_3 + \Delta \tilde C_3)^{*} - (19.62+1.81
  i)(\Delta C_5 + \Delta \tilde C_5)^{*}\nonumber \\ &&-
  (39.11 + 3.31 i )(\Delta C_7 + \Delta \tilde C_7)^{*} - (48.28 + 4.12
  i)(\Delta C_9 + \Delta \tilde C_9)^{*}]\nonumber \\ &&/[ 1 + (0.25+0.07 i
  )(\Delta C_3 + \Delta \tilde C_3) - ( 19.28+2.79 i)(\Delta C_5 +
  \Delta \tilde C_5)\nonumber \\ &&- (38.46 + 5.28 i
  )(\Delta C_7 + \Delta \tilde C_7) - ( 47.48 + 6.55 i)(\Delta C_9 +
  \Delta \tilde C_9)]~,  \nonumber 
  \end{eqnarray} 
\begin{eqnarray} 
\lambda_{\omega K_S} &=&  (-0.71 + 0.70 i)  \label{412}\\ \nn && [1  +
  (90.48+13.54 i )(\Delta C_3 + \Delta \tilde C_3)^{*} + (85.24+12.50
  i)(\Delta C_5 + \Delta \tilde C_5)^{*}\nonumber \\ &&+
  (32.21+4.80i )(\Delta C_7 + \Delta \tilde C_7)^{*} + (19.07 + 2.79
  i)(\Delta C_9 + \Delta \tilde C_9)^{*}]\nonumber \\ &&/[ 1+ (90.01+13.29 i
  )(\Delta C_3 + \Delta \tilde C_3) + ( 84.80+12.26 i)(\Delta C_5 +
  \Delta \tilde C_5)\nonumber \\ && + (32.04 + 4.71 i
  )(\Delta C_7 + \Delta \tilde C_7) + (18.97 + 2.74 i)(\Delta C_9 +
  \Delta \tilde C_9)],  \nonumber 
\end{eqnarray} 
\begin{eqnarray} 
\lambda_{f^0 K_S} &=&  (-0.70 + 0.70 i)  \label{413}\\ \nn && [1  +
  (1.02+0.42i )(\Delta C_3 + \Delta \tilde C_3)^{*} - (1.67+0.97
  i)(\Delta C_5 + \Delta \tilde C_5)^{*}\nonumber \\ &&+
  (3.19 + 0.93 i )(\Delta C_7 + \Delta \tilde C_7)^{*} - (0.12 + 0.15
  i)(\Delta C_9 + \Delta \tilde C_9)^{*}] \nonumber \\ && /[1 + (1.01+0.40 i
  )(\Delta C_3 + \Delta \tilde C_3) - ( 1.65+0.95 i)(\Delta C_5 +
  \Delta \tilde C_5)\nonumber \\ &&+ (3.16 + 0.90 i
  )(\Delta C_7 + \Delta \tilde C_7) - ( 0.12 + 0.15 i)(\Delta C_9 +
  \Delta \tilde C_9)]~.  \nonumber 
\end{eqnarray} 
These results are more general than those of ~\cite{Barger:2009eq,Barger:2009qs}, as they include the $Z'$ contributions to both the QCD and electroweak penguins.  
At the leading order, the deviations for ${\mathcal C}_{f_{CP}}$ and ${\mathcal S}_{f_{CP}}$ from their SM predictions are a linear combination of these two classes of $Z'$ contributions. This discussion is independent of the details of the $U(1)'$ charges, so it can be applied to 
other family non-universal models as well; however, in the $U(1)_I$ model, the only non-trivial corrections are $\Delta \tilde C_3$ and $\Delta \tilde C_9$. 
\begin{figure}[ht]
\begin{center}
\includegraphics[width=0.60\textwidth]{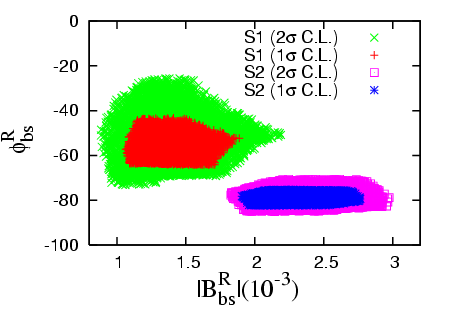}
\caption{Correlated constraints on $|B_{bs}^R|$ and $\phi_{bs}^R$.  Random values  for $C_{B_s}$ and $\phi_{B_s}^{\rm NP}$ from the experimentally allowed regions at different C.L. (see Table~\ref{table1}) are mapped to the $|B_{bs}^R|-\phi_{bs}^R$ plane using Eq.~(\ref{403}).}
\label{figure1}
\end{center}
\end{figure}


\begin{figure}[ht]
\begin{center}
\includegraphics[width=0.45\textwidth]{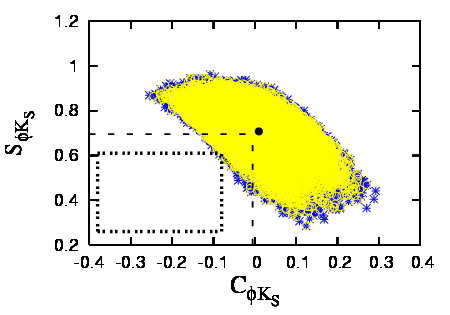}
\includegraphics[width=0.45\textwidth]{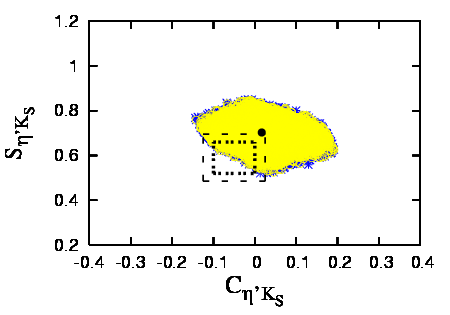}
\includegraphics[width=0.45\textwidth]{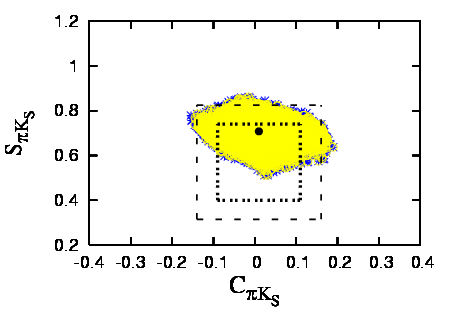}
\includegraphics[width=0.45\textwidth]{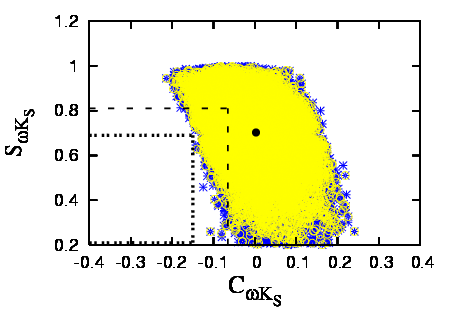}
\includegraphics[width=0.45\textwidth]{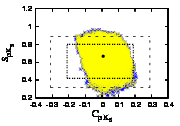}
\includegraphics[width=0.45\textwidth]{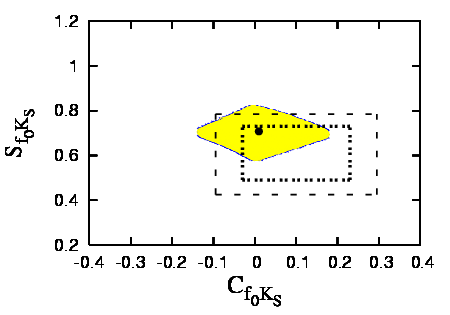}
\caption{
The NP contributions to  ${\mathcal C}_{(\phi, \eta', \rho, \omega, f_0)K_S}$ and ${\mathcal S}_{(\phi, \eta', \pi, \rho, \omega, f_0)K_S}$, with $|B_{bs}^R|$, $\phi_{bs}^R$ constrained by $B_s-\bar B_s$ mixing. The colors specify the C.L. that their inverse image points represent in Fig.~\ref{figure1} (yellow for $1 \sigma$ and blue for $2 \sigma$).  The boxes specify the allowed regions at 1$\sigma$ and $1.5\sigma$, and the dark points denote the SM limit. }
\label{figure2}
\end{center}
\end{figure}


\begin{figure}[ht]
\begin{center}
\includegraphics[width=0.45\textwidth]{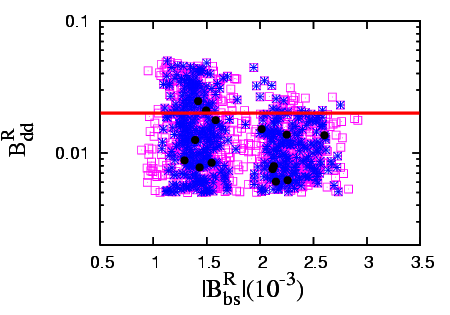}
\includegraphics[width=0.45\textwidth]{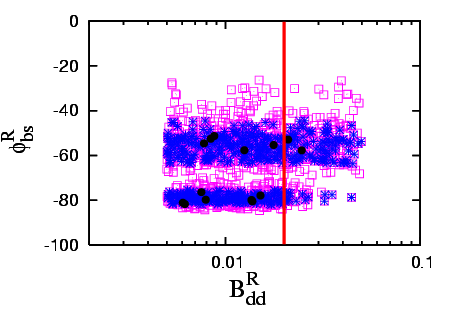}
\includegraphics[width=0.45\textwidth]{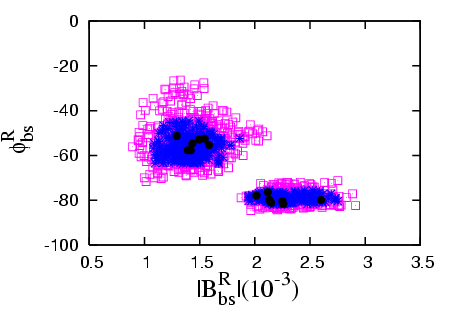}
\caption{The $|B_{bs}^{L,R}|$, $\phi_{bs}^{L,R}$ and $B_{dd}^R$ distributions, with values constrained by $B_s-\bar B_s$ mixing at $x \sigma$ C.L. and selected by  ${\mathcal C}_{(\phi, \eta', \pi, \rho, \omega, f_0)K_S}$ and ${\mathcal S}_{(\phi, \eta', \pi, \rho, \omega, f_0)K_S}$ at $y \sigma$ C.L.. Here $x=2.0$ and $y=1.7$ for the purple points, $x=1.0$ and $y=1.7$ for the blue points, $x=1.0$ and $y=1.4$ for the dark points. The red lines represent the vacuum considered in Section~\ref{spectrum}, in which the values of $|B_{bs}^R|$ and $\phi_{bs}^R$ are not fixed.}
\label{figure4}
\end{center}
\end{figure}
We now turn to a numerical analysis of the FCNC constraints with the $U(1)_I$ model, for which the three free parameters are $|B_{bs}^R|$, $\phi_{bs}^R$ and $B_{dd}^R$.  First, we consider $B_s-\bar{B}_s$ mixing, which involves two of these parameters, $|B_{bs}^R|$ and $\phi_{bs}^R$. The experimental constraints on these two parameters are illustrated in  Fig.~\ref{figure1}, where the various colors of the points specify the different confidence levels (C.L.) that the relevant $C_{B_s}$ and $\phi_{B_s}^{\rm NP}$ values represent.  There are two separate shaded regions in this figure. The left one corresponds to the $\phi_{B_s}^{\rm NP}$ solution ``S1'' and the right one corresponds to ``S2'' (see Tab.~\ref{table5}). $\phi_{bs}^R$ varies within the ranges $-80^\circ \sim -20^\circ$ and $-90^\circ \sim -70^\circ$ in the two regions, respectively. This is similar to what happens to $\phi_{bs}^L$ in the LL limit in~\cite{Barger:2009qs}, since the $\Delta \tilde C_3^{B_s}$ contributions to $C_{B_s} e^{2i\phi_{B_s}}$ in Eq.~(\ref{403}) are absent in both cases. In addition, to explain the observed discrepancy in $B_s-\bar B_s$ mixing from the SM prediction, $|B_{bs}^R|$ is required to be $\sim 10^{-3}$.  As discussed in~\cite{Barger:2009eq,Barger:2009qs}, there are two reasons for this feature. First, $C_{B_s}$ does not deviate significantly from its SM prediction (the anomaly in $B_s -\bar B_s$ mixing is mainly caused by the phase $\phi_{B_s}^{\rm NP}$). Second,  the corrections of a family non-universal $Z'$ arise at tree level, so only a small coupling is needed to explain this small deviation, according to Eq.~(\ref{403}). The smallness of $|B_{bs}^R|$ is consistent with our assumption of small fermion mixing angles, since $B_{bs}^R$ is proportional to them  (see Eq.~(\ref{457})) as well as to $g' M_{Z_1}/(g_Z M_{Z_2})$. The constraints from the branching ratio ${\rm Br}(B_s \to \mu^+ \mu^-)$ and ${\rm Br}(B_d \to K^{(*)}\mu^+\mu^-)$ can be easily satisfied due to the smallness of $|B_{bs}^R|$~\cite{Barger:2009qs}.

With the constrained values of $|B_{bs}^R|$ and $\phi_{bs}^R$ by $B_s-\bar B_s$ mixing, we illustrate the NP contributions to  ${\mathcal C}_{(\phi, \eta', \pi, \rho, \omega, f_0)K_S}$ and ${\mathcal S}_{(\phi, \eta', \pi, \rho, \omega, f_0)K_S}$ in Fig.~\ref{figure2}.  In this case, the third parameter ($B_{dd}^R$) is also involved. For the channel $B_d\to \pi K_S$, we take a strategy different from that used in~\cite{Barger:2009eq,Barger:2009qs}, in which it was assumed that  the NP enters the hadronic decays of neutral $B_d$ meson only through electroweak penguins.  In that case, the NP effects in the $B_d\to \pi K_S$ channel can be resolved into a factor $qe^{i\phi}$~\cite{Buras:2003dj}; the constraints on this factor from a $\chi^2$ fit of $B\to \pi K_S$ and $B \to \pi\pi$ data have been studied in~\cite{Fleischer:2008wb}. For our NUSSM model, the NP enters generically through QCD as well as electroweak penguins, and hence we treat this channel in the same way as the other $B_d$ decay channels. We also assume a $15\%$ uncertainty in the SM calculations for each of these modes and a $25\%$ uncertainty for the NP contributions.  Here $15\%$ is a typical uncertainty level for the hadronic matrix elements of the SM FC operators (see e.g.~\cite{Wirbel:1985ji}) that is needed to explain the experimental results for ${\mathcal C}_{\psi K_S}$ and ${\mathcal S}_{\psi K_S}$ in the SM~~\cite{Barger:2009eq,Barger:2009qs}. 
The difference of the uncertainty levels between the SM and NP calculations arises because the hadronic matrix elements of the FC operators in the SM are better understood than those of the NP operators. To see whether the anomalies in $B_s-\bar B_s$ mixing and the $B_d \to (\phi, \eta', \pi, \rho, \omega, f_0)K_S$  CP asymmetries can be simultaneously accommodated, we have carried out a correlated analysis within the $U(1)_I$ model.  The distributions of $|B_{bs}^R|$, $\phi_{bs}^R$ and $B_{dd}^R$ constrained at different C.L. are illustrated in Fig.~\ref{figure4}. Indeed, there exist parameter regions for which the tension between the observations and the SM predictions are greatly relaxed.

In Figs.~\ref{figure2} and~\ref{figure4}, we require $0.005 < |B_{dd}^R| < 0.5$. Given that 
$|B_{dd}^R| \approx g' M_{Z_1}/(2g_Z M_{Z_2})$, 
we immediately find that $1\, {\rm TeV} < M_{Z_2} < 10\, {\rm TeV}$ for $g' \simeq g_Z$. Here $B_{dd}^R$ can be positive or negative, since it resolves a minus sign from the degeneracy of two solutions in $B_{bs}^R$ that is specified by a $\pi$ phase difference~\cite{Barger:2009eq,Barger:2009qs}.   
The red lines in Fig.~\ref{figure4} represent the parameter region discussed in our numerical example in which $|B_{dd}^R| \approx 0.02$. Indeed, we see that the anomalies in the hadronic $B_d$ meson decays can be explained simultaneously, given the $B_{bs}^R$ values required to fit the $B_s\to \bar B_s$ mixing data.

\section{Discussion and Conclusions}
In this paper, we have discussed a class of family non-universal $U(1)^\prime$ models based on non-standard $E_6$ embeddings of the SM that interchange the standard roles of the two ${\bf 5^*}$ representations present in the fundamental ${\bf 27}$ representation of $E_6$ for the third family.  The NUSSM models in this class are simple and anomaly-free.  They are not full  $E_6$ grand unified theories, so  the $U(1)^\prime$ breaking can occur at the TeV scale, resulting in a TeV-scale $Z^\prime$ gauge boson that can mediate FCNC in the $b\to s$ transitions.  We analyzed a representative example of a NUSSM model (the $U(1)_I$ model), in which we described the low energy spectrum of the theory and determined the constraints on the family non-universal $Z^\prime$ couplings from the $B$ sector.  NUSSM models such as the $U(1)_I$ model are characterized by a rich spectrum of states with masses at the electroweak to TeV scale.  The $Z^\prime$-mediated FCNC in the $U(1)_I$ model can  easily accommodate the observed discrepancies in the $b\to s$ transitions.   Related observables such as $\tau\to \mu$ and $\tau \to e$ can also be studied in NUSSM models; we defer this to future work.

\section*{Acknowledgments}
We thank Carlos E.~M.~Wagner for helpful discussions and the Aspen Center for Physics for hospitality in the preparation of this work.  The work of L.~E.~is supported by the DOE grant No. DE-FG02-95ER40896 and the Wisconsin Alumni Research Foundation. The work of P.~L.~is supported by the IBM Einstein Fellowship and by the NSF grant PHY-0503584. The work of T.~L.~is supported by the Fermi-McCormick Fellowship and by the DOE grant No. DE-FG02- 90ER40560.

\newpage

\appendix

\renewcommand{\thesection}{Appendix \Alph{section}}

\setcounter{equation}{0}

\section{Tree-level Mass-squared Matrix for Charged Higgs Bosons}

\label{CH mass matrix}

\renewcommand{\theequation}{\Alph{section}.\arabic{equation}}

For charged Higgs bosons, the entries of its mass-squared matrix $M_{H^\pm}^2$ at tree level are given in the basis $\{h_d^-,h_u^{+*},H_d^-,H_u^{+*}\}$ by 
\begin{eqnarray}
(M_{H^\pm}^2)_{11} &=& -{{G^2}\over 4} \left(|v_2|^2 - |v_1|^2+|V_2|^2 - |V_1|^2\right)+ \frac{g_2^2}{2}(|v_2|^2+ |V_2|^2-|V_1|^2) \nonumber \\&& -\frac{g'^2}{4}\left(-|v_1|^2 + |s_1|^2 + 
|V_1|^2 -|s_2|^2\right)+(|\lambda_1|^2+|\lambda_2|^2)|s_1|^2+m_{h_d}^2~,~\,    \nonumber \\
(M_{H^\pm}^2)_{22} &=&  {{G^2}\over 4} \left(|v_2|^2 - |v_1|^2+|V_2|^2 - |V_1|^2\right)+\frac{g_2^2}{2} (|v_1|^2+|V_1|^2-|V_2|^2)\nonumber \\&&+|\lambda_1|^2|s_1|^2+|\lambda_3|^2|s_2|^2+m_{h_u}^2 ~,~\,  \nonumber \\
(M_{H^\pm}^2)_{33} &=& -{{G^2}\over 4} \left(|v_2|^2 - |v_1|^2+|V_2|^2 - |V_1|^2\right)+\frac{g_2^2}{2} (|v_2|^2+ |V_2|^2-|v_1|^2)\nonumber \\&& + \frac{g'^2}{4}\left(-|v_1|^2 + |s_1|^2 + 
|V_1|^2 -|s_2|^2\right)+(|\lambda_3|^2+|\lambda_4|^2)|s_2|^2+m_{H_d}^2 ~,~\,  \nonumber \\
(M_{H^\pm}^2)_{44} &=&  {{G^2}\over 4} \left(|v_2|^2 - |v_1|^2+|V_2|^2 - |V_1|^2\right)+ \frac{g_2^2}{2} (|v_1|^2+|V_1|^2-|v_2|^2)\nonumber \\&& +|\lambda_2|^2|s_1|^2+|\lambda_4|^2|s_2|^2+m_{H_u}^2 ~,~\,  \nonumber \\
(M_{H^\pm}^2)_{12} &=&(M_{H^\pm}^2)_{21}^* =\frac{g_2^2}{2} v_1^*v_2^* - \lambda_1(\lambda_1^*v_2^*v_1^*+\lambda_2^*V_2^*v_1^*+\lambda_5^* s_2^*)-A_{\lambda_1} \lambda_1 s_1  ~,~\,  \nonumber \\
(M_{H^\pm}^2)_{13} &=& (M_{H^\pm}^2)_{31}^* =\frac{g_2^2}{2} v_1^*V_1  + (\lambda_1\lambda_3^*+\lambda_2\lambda_4^*) s_1 s_2^*  ~,~\,  \nonumber \\
(M_{H^\pm}^2)_{14} &=& (M_{H^\pm}^2)_{41}^* =\frac{g_2^2}{2} v_1^*V_2^*  - \lambda_2(\lambda_1^*v_2^*v_1^*+\lambda_2^*V_2^*v_1^*+\lambda_5^* s_2^*)-A_{\lambda_2} \lambda_2 s_1 ~,~\,  \nonumber \\
(M_{H^\pm}^2)_{23} &=& (M_{H^\pm}^2)_{32}^* =\frac{g_2^2}{2} v_2V_1  - \lambda_3^*(\lambda_3h_u^{0}V_1+\lambda_4V_2V_1+\lambda_5 s_1) -A_{\lambda_3}^* \lambda_3^* s_2^* ~,~\,  \nonumber\\
(M_{H^\pm}^2)_{24} &=& (M_{H^\pm}^2)_{42}^* =  \frac{g_2^2}{2} v_2V_2^* + \lambda_1^*\lambda_2 |s_1|^2  + \lambda_3^*\lambda_4 |s_2|^2 ~,~\,   \nonumber \\
(M_{H^\pm}^2)_{34} &=& (M_{H^\pm}^2)_{43}^* = \frac{g_2^2}{2} V_1^*V_2^* -\lambda_4(\lambda_3^*v_2^*V_1^*+\lambda_4^*V_2^*V_1^*+\lambda_5^* s_1^*)  -A_{\lambda_4} \lambda_4 s_2 ~.~\,  \nonumber 
\end{eqnarray}
These entries can be applied to both cases with and without CP violation.

\renewcommand{\thesection}{Appendix \Alph{section}}

\setcounter{equation}{0}

\section{Parameters}

\label{Parameters}

\renewcommand{\theequation}{\Alph{section}.\arabic{equation}}

The parameters used in our numerical analysis are summarized below:

\bigskip

{\bf (1) QCD and EW Parameters}

\begin{center}
$G_F = 1.16639 \times 10^{-5}$ GeV$^{-2}$, 
\hspace{20mm}$\Lambda_{\overline{MS}}^{(5)} = 225$ MeV, \\
$M_W = 80.42$ GeV, \hspace{20mm} $\sin^2\theta_W = 0.23$, \\
$\eta_{2B} = 0.55$, \hspace{20mm} $J_5 = 1.627$, \\  
$\alpha_s(M_Z) = 0.118$,\hspace{20mm} $\alpha_{em} = 1/128$, \\
$\lambda = 0.2252$, \hspace{20mm} $A = 0.8117$, \\
$\bar{\rho} = 0.145$, \hspace{20mm} $\bar{\eta} = 0.339$, \\
$R_b = \sqrt{\rho^2 + \eta^2} = 0.378$. \\
\end{center}

\bigskip

{\bf (2) Masses, Decay Constants, Hadronic Form Factors and Lifetimes}
\begin{center}
$M_{{\pi}^{\pm}} =0.139$ GeV, \hspace{20mm} $M_{{\pi}^{0}} = 0.135$ GeV,\\
$M_{K} = 0.498$ GeV, \hspace{20mm} $M_{B} = 5.279$ GeV, \\
$M_\phi = 1.02$ GeV, \hspace{20mm} $M_\psi = 2.097$ GeV, \\
$M_{\eta^\prime} = 0.958$ GeV, \hspace{20mm} $M_{\omega} = 0.783$
GeV, \\
$M_{\rho} = 0.776$ GeV, \hspace{20mm} $M_\eta = 0.548$ GeV \\
$M_{f^0} = 0.980$ GeV, \\
$X_{\eta} = 0.57$, \hspace{20mm} $Y_{\eta} = 0.82$, \\
$m_u (\mu = 4.2 ~{\rm GeV}) = 1.86$ MeV, \hspace{20mm} $m_d (\mu = 4.2
~{\rm GeV})= 4.22$ MeV, \\ 
$m_s (\mu = 4.2 ~{\rm GeV}) = 80$ MeV, \hspace{20mm} $m_c (\mu = 4.2
~{\rm GeV}) = 0.901$ GeV, \\ 
$m_b(\mu = 4.2 ~{\rm GeV}) = 4.2$ GeV, \hspace{20mm} $m_t (\mu = M_Z)
= 171.7$ GeV,\\ 
$f_{\phi} = 237$ MeV, \hspace{20mm} $f_{B} = 190$ MeV,\\
$f_{\pi} = 130$ MeV, \hspace{20mm} $f_{K} = 160$ MeV,\\
$f_{\psi} = 410$ MeV, \hspace{20mm} $f_{\omega} = 200$ MeV, \\
$f_\rho = 209$ MeV, \hspace{20mm} $f_{f^0} = 180$ MeV,\\
$F_0^{B\pi} (0) = 0.330$, \hspace{20mm} $F_0^{BK}
(0) = 0.391$, \\ 
$F_1^{BK} (0) = 0.379$, \hspace{20mm} $A_0^{B\omega} (0) =
0.280$, \\  
$F_0^{B f} (0) = 0.250$, \hspace{20mm} $F_0^{f K} (0) = 0.030$, \\
$A_0^{B\rho} = 0.280$, \hspace{20mm} $f_{B_s} \sqrt{{\hat B}_{B_s}} =
0.262$ \\ 
$\tau_{B^0}=1.530$ ps, \hspace{20mm} $\tau_{B^-}=1.65$ ps, \\
$M_{B_s} = 5.37$ GeV, \hspace{20mm} $\tau_{B_s} = 1.47$ ps, 
\end{center}

\newpage

\bibliographystyle{prsty}

\end{document}